\documentclass[reprint,amsmath,amssymb,aip,jcp,author-numerical]{revtex4-1}
\usepackage{graphics}
\usepackage{graphicx}
\usepackage{dcolumn}
\usepackage{bm}
\usepackage{natbib}
\usepackage{hyperref}
\usepackage[usenames,dvipsnames,svgnames,table]{xcolor}

\begin{document}

\title{Gaussian field model for polar fluids as a function of density and polarization: toward a model for water.}
\author{H. Berthoumieux}

\affiliation{CNRS, UMR 7600, LPTMC, F-75005, Paris, France}
\affiliation{Sorbonne Universit\'es, UPMC Univ Paris 06, UMR 7600, LPTMC, F-75005, Paris, France}
\begin{abstract}
This work is concerned with a simple model for a polar fluid, a Gaussian field model based on the excess density and on the polarization. It is a convenient framework to implement the dielectric properties of correlated liquids that stem from nanometric correlations between molecules. It allows to study the effects of coupling terms between density and polarization on the structure of the fluid. Despite the simplicity of such a model, it can capture some interesting features of the response functions of water such as the quasi-resonant longitudinal dielectric susceptibility or the presence of two maxima in the structure factor. Explicit models of water generate extremely high computational cost and implicit models sometimes fail to treat properly the electrostatic interactions. A Gaussian field theory could  therefore be an interesting alternative to describe water. 
\end{abstract}
\maketitle
\section{Introduction}

Water is the most abundant fluid on Earth and a prerequisite for life on this planet.
It is an associated fluid structured by H-bond networks. The strength of these interactions and the cavities created by the spacial organization that they generate give rise to many anomalous properties of water \cite{cabane}. Water as a solvent plays an essential role to assist a broad range of nanometric machines in their functions, proteins\cite{fogarty2013, bellissent-funel2016} or nanocapacitors \cite{limmer2011} to name a few. The 'perfect' model of water capturing enough microscopic details to keep track of its molecular structure but coarse-grained enough to reach an affordable computational cost is an active domain of research.


Atomistic models for water give acces to a large amount of molecular detail \cite{mark2001} but these models imply exorbitant computational times and alternatives have to be envisaged when one studies large systems or systems in which the solvent is not pure, like electrolytes \cite{chen2016,pluharova2017}.  On the other hand, implicit water models consisting in a mean force applied on objects embedded in the solvent can be significantly faster but they neglect length-scales originating in the molecular structure of the solvent \cite{kleinjung2014}. 

Continuous models describing water and using density as an order parameter such as gaussian models or functional models (DFT) have brought a huge contribution to the field of the physics of liquids, clarifying the structural response of the fluid to microscopic and macroscopic inclusions by taking into account short and long range density fluctuations \cite{lum1999,willard2010}.  However, these models that focus on density effects do not properly reproduce electrostatic interactions in water. 

It is worth recalling that water possesses peculiar dielectric properties that result from the 
hydrogen-bond network structuring the fluid and correlating water dipole orientations on a nanometric lengthscale. The dielectric susceptibility is not constant but exhibits a quasi-resonance in Fourier space for a wave number corresponding to the inverse H-bond length. Neutron scattering and molecular dynamics simulations have given access to its precise shape \cite{bopp1996}.  Continuous theories can capture these properties emerging from molecular correlations. They are often referred to as 'nonlocal' \cite{kornyshev1976,kornyshev1981,fedorov2007,paillusson2010,ren2017}. The medium is described by a field, often the polarization, and associated with a Landau-Ginzburg Hamiltonian the expansion of which is chosen to reproduce the experimentally measured susceptibility. A model capturing key properties of water dielectric susceptibility was developed \cite{maggs2006} and used to show that short-range correlations between water molecules strongly affect the interaction between charged inclusions embedded in the fluid \cite{rottler2009} and the electrostatic attraction between macroscopic neutral objects immersed in water \cite{berthoumieux2015}. 

Recently, a classical density functional theory (MDFT) using the density and the polarization was proposed to describe water \cite{jeanmairet2013,jeanmairetthese,cageat2017}.  A few physical quantities such as the structure factor and the dielectric susceptibilities are used as inputs of the MDFT. This numerical method gives satisfactory results for solvation energy and hydration profile of any solute with a significantly reduced computational cost. 

In this  work we propose a first step toward a tractable analytical framework to model water at an intermediate length-scale, {\it i.e.} including nanometric details coming out of the molecular structure of the fluid without keeping track of all the atomistic details.  The article is organized as follows. In the first part, we propose a Gaussian field model for a polar liquid which is described by a density and a polarization field. We use a Landau-Ginzburg expansion to obtain an Hamiltonian that gives rise to nanometric correlation lengths for density and polarization.  
In a second part, we perturb the previous model by adding a coupling term between the two order parameters. We study the form of the coupling, derive the susceptibilities and the Green functions of the system. In the third part, we use this extended Hamiltonian to reproduce qualitatively the susceptibilities of water.  We derive the radial distribution function and the response function of the medium to an impurity. We show that this approach can capture some of  the structural properties of water. The last part is devoted to the discussion and the conclusion. 

\section{Two order parameter Gaussian model for a polar fluid}
 We start by considering a simple model of fluid, a Gaussian model associated with a scalar field, the density $\rho(r)$, and a vectorial field, the polarization ${\bf P}(r)$. 
 The Hamiltonian of the system is thus  characterized by two susceptibilities and can be written as
\begin{eqnarray}
\label{Hgenenc}
\mathcal{H}[\delta \rho, {\bf P}]&=&\frac{k_bT}{2\rho_0}\int d^3r d^3r' \delta \rho (r) \chi^{-1}_{\rho,0}(r,r') \delta \rho (r')\nonumber\\
&+&\frac{1}{2\epsilon_0}\int d^3r d^3r'{\bf P}(r)\cdot {\bf \chi}^{-1}_{P,0}(r,r')\cdot {\bf P}(r'),
\end{eqnarray}
where $\delta\rho(r)$ is the deviation of the fluid density from its mean $\rho_0=\rho(r)-\delta\rho(r)$. The polarization  ${\bf P}(r)$ is defined as the average of the dipolar moments on a mesoscopic volume.
$k_B$ is the Boltzmann constant and $T$ is the temperature, $\rho_0$ is the density of the fluid and taken in this paper to be $\rho_0=0.033 \AA^{-3}$ (the density of water in normal conditions), and $\epsilon_0$ the permittivity of the vacuum. $\chi_{\rho,0}$ and $\chi_{P,0}$ are respectively the density susceptibility and the tensorial dielectric susceptibility. 
The unperturbed fluid is characterized by a homogeneous density equal to $\rho_0$ and a vanishing polarization.

The first term of the Hamiltonian corresponds to the energy of excess density \cite{chandler1993}, the second term is the dielectric energy of the medium.  
The variance of the fields $\delta\rho(r)$ and ${\bf P(r)}$ are linked to the susceptibilities introduced in Eq. (\ref{Hgenenc}) by the relations:
\begin{eqnarray}
\label{densityvariance}
\langle \delta\rho(q)\delta\rho(-q) \rangle &=& \rho_0\chi_{\rho,0}(q), \nonumber\\
\langle  P_i(q) P_j(q) \rangle &=& k_bT\epsilon_0 \chi_{P,0 (i,j)}(q).
\end{eqnarray}
The structure factor $S(q)$ of such a fluid is equal to the density susceptibility $\chi_{\rho,0}(r)$\cite{sergilevskyl2017}.


In a first step, we propose a Landau-Ginzburg expansion for the density and polarization terms including the first and the second spacial derivatives of the fields.
The Hamiltonian can then be written as:
\begin{widetext}
\begin{eqnarray}
\label{Hnc}
\mathcal{H}[\delta\rho, {\bf P}]&=&\frac{k_bT}{2\rho_0}\int d^3r \Big[ K_{\rho}\delta\rho(r)^2+\lambda\left(\nabla \delta\rho(r)\right)^2
+\nu \left(\nabla \cdot \nabla\delta\rho(r)\right)^2 \Big]\nonumber\\
 &+&\frac{1}{2\epsilon_0}\int d^3r\
\Big [ K{\bf P}^2(r)+\kappa_l({\nabla}\cdot{\bf P}^(r))^2+\alpha {(\nabla
                                     (\nabla \cdot {\bf
                                     P}(r))  )}^2 \Big ]  
                                  + \frac{1}{2\epsilon_0} \int d^3r d^3r' \frac{\nabla \cdot {\bf P}(r) \nabla \cdot {\bf P}(r')}{4\pi| r-r'|}. 
\end{eqnarray}
\end{widetext}
Note that the dielectric energy contains a local contribution due to the short-range interactions between dipoles  and a long-range part coming from the Coulomb interactions.  The dielectric susceptibility tensor is characterized by a longitudinal and a transverse scalar susceptibility such that $\chi_{P,0 (i,j)}(q)=\chi_{\parallel}(q)q_iq_j/q^2+\chi_\perp(q)(\delta_{ij}-q_iq_j/q^2)$. The transverse susceptibility associated with this Hamiltonian is purely local and equal to $\chi_{\perp,0}(q)=1/K$. In the following, we will consider only the longitudinal susceptibility, which is equivalent to consider only the longitudinal contribution of the field {\bf P} such that $\nabla \times {\bf P}=0$.

The corresponding density and dielectric susceptibility tensors in Fourier space are equal to
\begin{eqnarray}
\label{SF0}
\chi_{\rho,0}(q)&=&\frac{1}{K_{\rho}+\lambda q^2+\nu q^4}\\
\label{chi}
\chi_{P,0(i,j)}(q)&=&\chi_{P,0}(q)\frac{q_iq_j}{q^2},\nonumber\\
  {\rm with } \quad \chi_{P,0}(q)&=&\frac{1}{1+K+\kappa_l q^2+\alpha q^4}.
\end{eqnarray}
The Landau-Ginzburg expansion to order $q^4$ can give rise to susceptibilities exhibiting a maximum in $q$-space if one imposes negative values for $\lambda$ and $\kappa_l$ and positives values for $\alpha$ and $\nu$ which ensure the stability of the system. 
The values for $q=0$ of the susceptibilities $\chi_{P,0}(0)$ and $\chi_{\rho,0}(0)$ are linked to the macroscopic properties of the fluid, {\it i.e.} to the permitivity and the compressibility through the relations $\chi_{P,0}(0)=\frac{1}{1-\epsilon}$ and  $\chi_{\rho,0}(0)=\chi_T/\chi_T^0$, where $\chi_T$ is the compressibility of fluid and $\chi_T^0=1/k_bT\rho_0$ is the compressibility of a perfect gas of density $\rho_0$. 
We choose the values of $K$ and $K_{\rho}$ such that the macroscopic properties of the fluid described by Eq. (\ref{Hnc}) correspond to SPC/E water properties \cite{reddy1989}. The values of $(\lambda, \nu)$ and $(\kappa_l, \alpha)$ are chosen so that the value and the position of the maxima of $\chi_{\rho,0}(q)$ and $\chi_{P,0}(q)$ fit the ones SPC/E water \cite{bopp1996,jeanmairet2013,jeanmairetthese}.
\begin{figure*}
\includegraphics[scale=0.4]{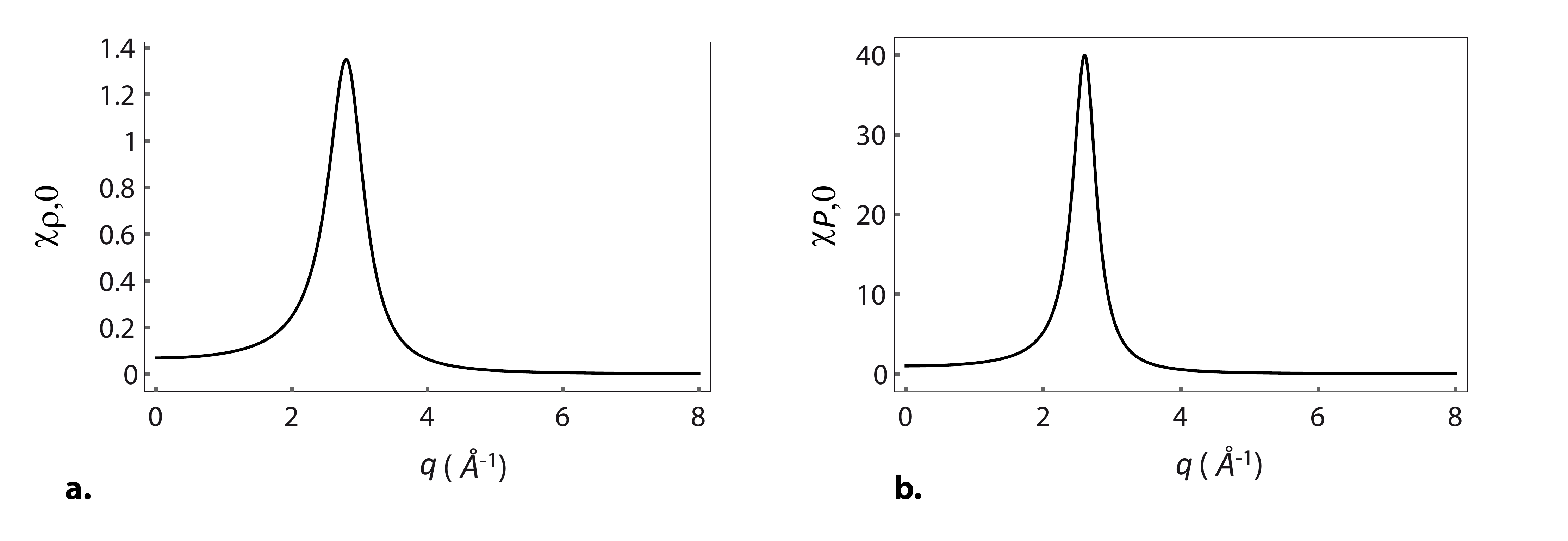}
\caption{{\bf a}. Density susceptibility $\chi_{\rho,0}(q)$ as a function of the wavevector.
The expression of $\chi_{\rho,0}(q)$ is given in Eq. (\ref{SF0}) and plotted for $K_{\rho}=14.44$, $\lambda=-3.50$ \AA$^{-2}$ and $\nu=0.22$ \AA$^{-4}$. {\bf b}.
Longitudinal susceptibility $\chi_{P,0}(q)$ as a function of the wavevector. The expression of $\chi_{P,0}(q)$ is given in Eq. (\ref{chi}) and is plotted for $K=1/70$, $\kappa_l=-0.29$ \AA$^{-2}$, $\alpha=0.02$ \AA$^{-4}$. }
\end{figure*}

The density correlation function in the real space is obtained by Fourier transform of Eq. (\ref{densityvariance}) and is an oscillating function in an exponentially decaying envelope, 
\begin{equation}
\label{rhorho0}
\langle \delta \rho (r) \delta \rho (0) \rangle =\rho_0\frac{(1+R_\rho^2)^2}{8\pi K_{\rho}R_{\rho}\lambda_{e,\rho}^2r}e^{-r/\lambda_{e,\rho}}\sin(\frac{r}{\lambda_{o,\rho}}),
\end{equation}
with $R_{\rho}=\lambda_{e,\rho}/\lambda_{o,\rho}$, $\lambda_{o,\rho}=\sqrt{2}/q_{0,\rho}\sqrt{1/\sqrt{\zeta_{\rho}}+1}$, $\lambda_{e,\rho}=\sqrt{2}/q_{0,\rho}\sqrt{1/\sqrt{\zeta_{\rho}}-1}$, $\zeta_{\rho}=\nu q_{0,\rho}^4/K_{\rho}$. For the chosen set of parameters, the decay length $\lambda_{e,\rho}$ and the oscillating length $2 \pi \lambda_{o,\rho}$ are equal respectively to 0.35 nm and 0.22 nm. The polarization correlations are discussed in Appendix A.

The density correlation function is linked to the well characterized pair distribution function through the relation
\begin{equation}
\label{raddist}
g(r)=1+\frac{\langle \delta \rho (r) \delta \rho (0) \rangle}{\rho_0^2}.
\end{equation}  
The radial distribution function $g(r)$ associated with the Hamiltonian given in Eq. (\ref{Hnc}) is represented in Figure 2. The first peak, located around 3 \AA, corresponds to the first solvation shell of a reference molecule. The function decreases slowly in $r$ with $e^{-r/\lambda_{e,\rho}}/r$  and the oscillations due to the layers of ordered molecules are visible over a range of 1 nm. This is much larger than what is observed in liquid phase and in particular in water where the envelope of the radial distribution function decays more rapidly and vanishes after two oscillations\cite{sorenson2000}. Indeed, the density susceptibility obtained from the Landau-Ginzburg expansion in $q^4$ presents a narrow maximum which corresponds to a long-range correlated fluid.
\begin{figure}
\includegraphics[scale=0.4]{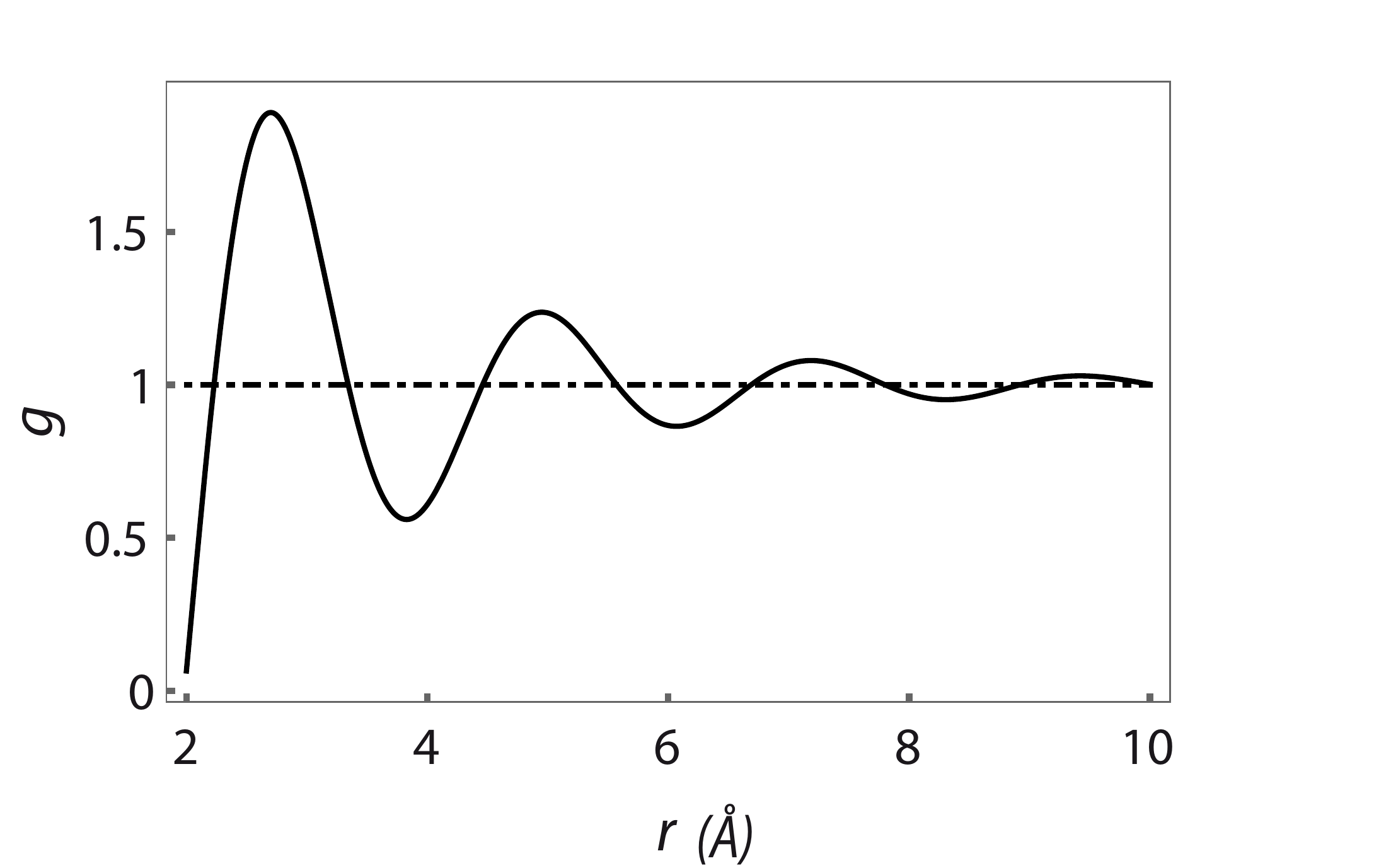}
\caption{The radial distribution $g(r)$ as a function of the distance. The expression of $g(r)$ is given in Eq. (\ref{raddist}) and plotted for parameter values given in Fig. 1.}
\end{figure}

The local density of structured liquids such as water is closely linked to the spacial organization of the molecular dipoles \cite{soper2000}. 
A coarse-grained desciption of a fluid associated with two coupled parameters could take into accound the interplay between structure and density. A Landau-Ginzburg Hamiltonian, function of $\rho$ and {\bf$ P$}, expanded to the second spacial derivative and generating a Lorentzian dielectric susceptibility was introduced by Kornyshev and coworkers \cite{kornyshev1997}. It was shown that a coupling between $\rho$ and {\bf $P$} generates a maximum in the dielectric susceptibility. In the next two sections, we study the effect of the coupling in the case of the model given by Eq. (\ref{Hnc}).

\section{Presence of a coupling term between density and polarization}

We add a coupling term between density and polarization to the Hamiltonian given in Eq. (\ref{Hnc}) and study its effects on the response functions of the system. Its form is dictated by symmetry arguments requiring that the coupling term is a scalar. A spacial derivative of one of the fields has to be introduced and a supplementary length scale will be present in the problem. We consider the gradient term and neglect higher derivative terms and add to the energy given in Eq. (\ref{Hnc}) an extra contribution   $c\frac{k_bT}{\mu_0\rho_0}\int d^3r\nabla\delta \rho (r) \cdot {\bf P}(r)$. 
The coupling constant $c$ is homogeneous to a length. Using partial integration, the coupling term can be written as $-c\frac{k_bT}{\mu_0\rho_0}\int d^3r\delta \rho (r)  \nabla\cdot {\bf P}=-c\frac{k_bT}{\mu_0\rho_0}\int d^3r\delta \rho (r) \cdot \rho_b(r)$, where $\rho_b$ is the local bound charge density. 

The Hamiltonian of the system characterized by the two coupled order parameters $(\delta \rho, { \bf P} )$  is now equal to:
\begin{eqnarray}
\label{Hcoupled}
\mathcal{H}_c[\delta\rho,{\bf P}]&=&\frac{k_bT}{2\rho_0}\int d^3q \delta \rho(q) \chi_{\rho,0}^{-1} \delta \rho (-q) \nonumber\\ &+&\frac{1}{2\epsilon_0}\int d^3q {\bf P}(q)\chi_{P,0}^{-1}{\bf P}(-q)\nonumber\\ &-&c \frac{i k_bT}{\mu_0 \rho_0}\int d^3 q{\bf q} {\bf P}(q)\delta\rho (-q)\nonumber\\
&=&\frac{1}{2}\int d^3q \left(\delta \rho(q), {\bf P}(q) \right)\chi^{-1}_{c}(q)\left(\begin{array}{c} \delta \rho(-q) \\ {\bf P}(-q)\end{array}\right).
\end{eqnarray}
The kernel $\chi_c^{-1}(q)$ is a 4x4 matrix that is given in Appendix B.
The density and the polarization susceptibilities are the diagonal terms of the inverse of this matrix and are equal to  
\begin{eqnarray}
\label{chirhoc}
\chi_{\rho,c}(q)&=&\frac{\chi_{P,0}^{-1}(q)}{\chi_{P,0}^{-1}(q)\chi_{\rho,0}^{-1}(q)-c^2 q^2\frac{k_bT\epsilon_0}{\mu_0^2\rho_0}},\\
\label{chipc}
{\bf \chi}_{(P,c)ij}&=&\chi_{P,c}(q)\frac{q_iq_j}{q^2} \nonumber\\
{\rm with} \quad \chi_{P,c}(q)&=&\frac{\chi_{\rho,0}^{-1}(q)}{\chi_{P,0}^{-1}(q)\chi_{\rho,0}^{-1}(q)-c^2 q^2\frac{k_bT\epsilon_0}{\mu_0^2\rho_0}}, 
\end{eqnarray}
 where $\chi_{\rho,0}(q)$ and $\chi_{P,0}(q)$ are the uncoupled susceptibilities given in Eqs. (\ref{SF0},\ref{chi}).
For a vanishing coupling term, we obtain the expressions given in Eqs (\ref{SF0},\ref{chi}).
The system is now characterized by a cross susceptibility $\chi_{\rho,P}(q)$ proportional to the density-charge density correlations that we define following Borgis and coworkers \cite{jeanmairet2016} as $\chi_{\rho,P}(q)=\left(\frac{\rho_0}{k_bT\epsilon_0}\right)^{1/2}\frac{\langle \rho_c (q) \delta \rho(-q)\rangle}{q}$, 
$\rho_c(q)=i {\bf q }{\bf P}(q)$ being the bound charge density. 
Using the correlation between the polarization and the density derived from Eq. (\ref{Hcoupled}), $\langle \delta\rho(q) P_i(-q) \rangle =icq_ik_bT\frac{\epsilon_0}{\mu_0}\frac{1}{\chi^{-1}_{\rho,0}(q)\chi_{p,0}^{-1}(q)-c^2q^2\frac{k_bT\epsilon_0}{\mu_0^2\rho_0}}$, we derive the expression of the cross susceptibility,
\begin{equation}
\label{chirhop}
\chi_{\rho,P}(q)=\left(\frac{k_bT \epsilon_0}{\rho_0}\right)^{1/2}\frac{cq}{\mu_0}\frac{1}{\chi^{-1}_{\rho,0}\chi_{p,0}^{-1}-c^2q^2\frac{k_bT\epsilon_0}{\mu_0^2\rho_0}}.
\end{equation}

\begin{figure*}
\includegraphics[scale=0.3]{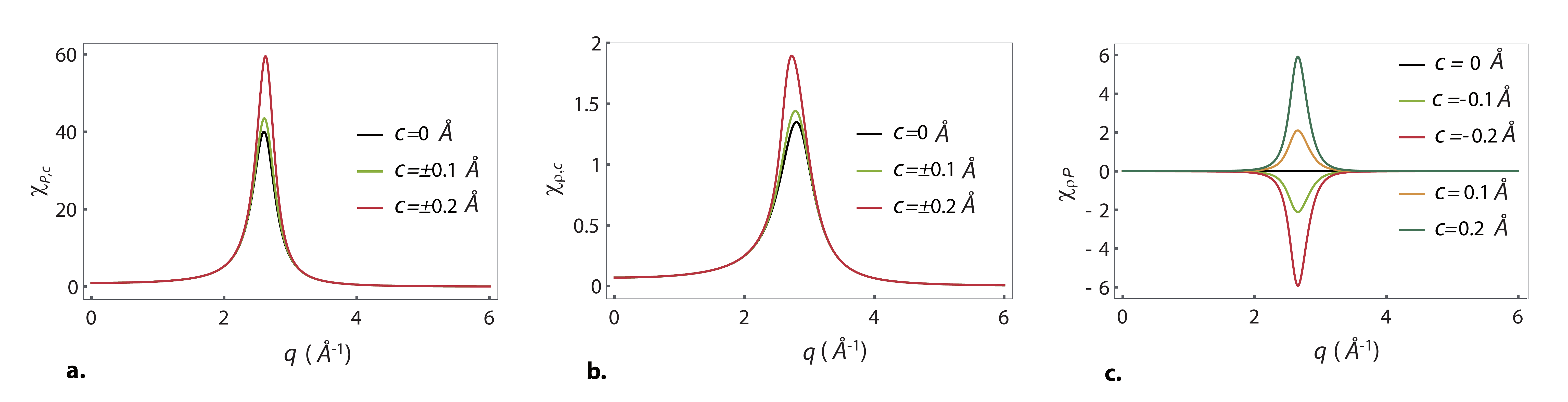}
\caption{{\bf a}. Density susceptibility $\chi_{\rho,c}(q)$ given in Eq. (\ref{chirhoc}) as a function of $q$ for different values of the coupling constant $c$.
{\bf b}. Longitudinal susceptibility $\chi_{P,c}(q)$ given in Eq. (\ref{chipc}) as a function of $q$ for different values of $c$.
{\bf c}. Longitudinal susceptibility $\chi_{\rho,P}(q)$ given in Eq. (\ref{chirhop}) as a function of $q$ for different values of $c$.
The functions are plotted for parameter values given in Fig. 1 }
\end{figure*}

$\chi_{\rho,c}(q)$, $\chi_{P,c}(q)$, and $\chi_{\rho,P}(q)$  are plotted in Figure 3 for the set of parameters given in Fig. 1 and increasing values of the coupling constant $c$. 
The functions $\chi_{\rho,c}(q)$ and $\chi_{P,c}(q)$ are functions of $c^2$ and do not depend on the sign of the coupling constant. The cross-susceptibility $\chi_{\rho,P}(q)$ is proportional to $c$ and presents a peak which is sharper than the peak of the density susceptibility but less sharp than the peak associated with $\chi_{P,c}(q)$. The coupling effectively broadens the peaks of the susceptibilities and displaces the maxima toward small $q$, respectively large $q$, for the density, respectively for the polarization. The susceptibilities diverge for a large value of $c$ showing that a strong coupling between ${\bf P}$ and $\delta \rho$ leads to an unstable system. A medium associated with the parameter values given in Fig. 1 is stable for $c< -0.33$ \AA. 

To gain physical insight into the expressions of the susceptibilities given in Eqs. (\ref{chirhoc}-\ref{chipc}), we consider the Green functions that control the response of an homogeneous medium characterized by the Hamiltonian Eq. (\ref{Hcoupled}) to a perturbation in density and in polarization. The Green functions are derived in Appendix C. A perturbation in density induces a response in density given by $G_{\rho}(r)$  and in polarization given by  $G_{\rho,P}$. A polarization of the medium induces a response in polarization and in density that are given respectively by the Green functions $G_{P (i,j)}(r)=T_c(r)(\delta_{ij}-r_ir_j/r^2)+L_c(r)r_ir_j/r^2$ and $G_{\rho,P}(r)$.
These functions are plotted in Fig. 4 for a vanishing coupling, $c=0$ and a negative coupling $c=$-0.2 \AA.
\begin{figure*}
\includegraphics[scale=0.4]{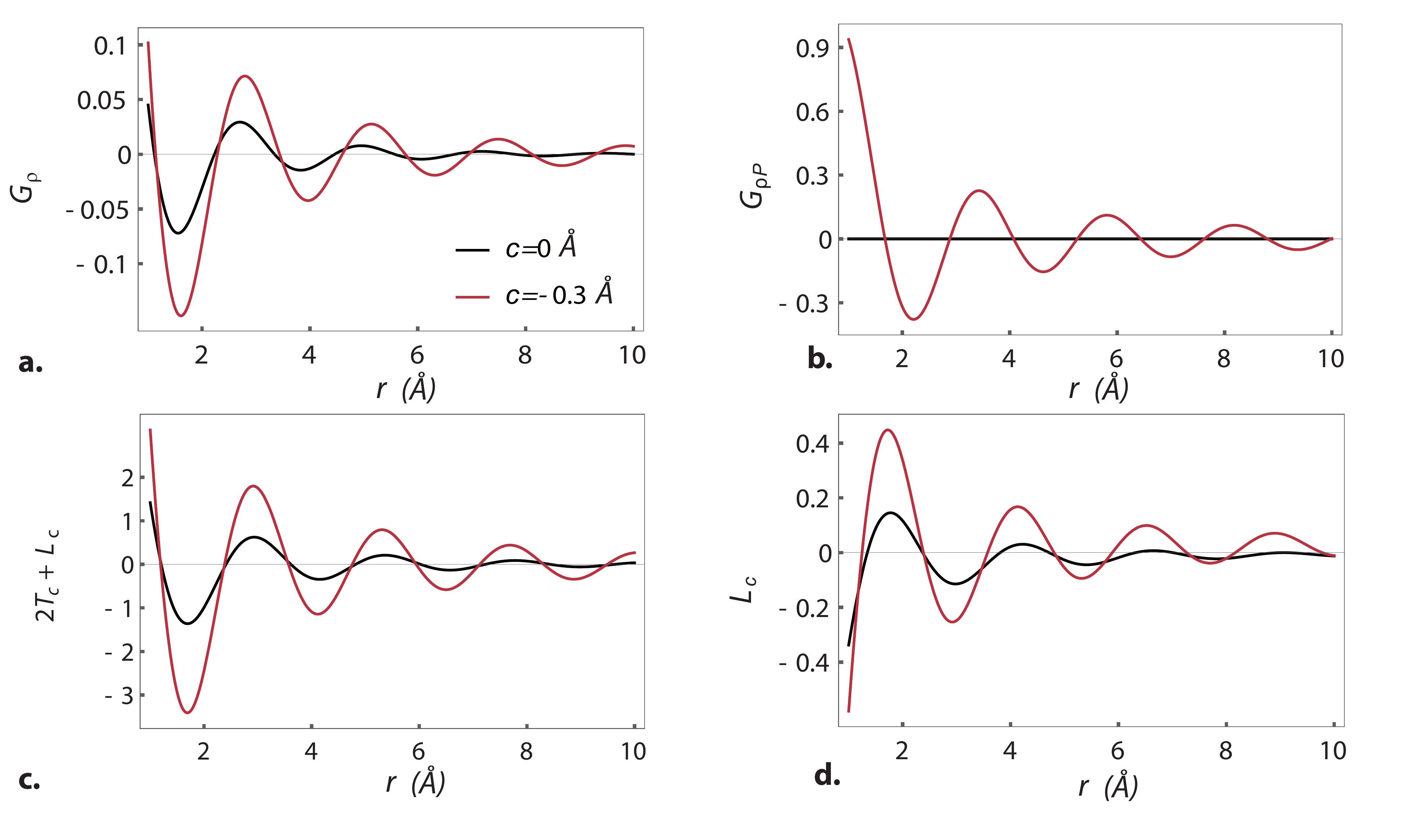}
\caption{Green functions of the system. {\bf a}. $G_{\rho}(r)$ is given in Eq. (\ref{Grho}).  {\bf b}. $G_{\rho,P}(r)$ is given in Eq. (\ref{Grhop}), {\bf c}. $2 T_c(r)+L_c(r)$ is given in Eq. (\ref{TcLc}). {\bf d}. $L_c(r)$  is given in Eq. (\ref{Lc}). The functions are plotted for the values of parameters given in Fig. 1 }
\end{figure*}
They are the sum of two oscillating functions in exponentially decaying envelopes, each contribution being associated with a decaying length that defines the range of the response. In the absence of coupling, the range of $G_{\rho}(r)$, $G_{P}(r)$ respectively,  is equal to 0.31 nm, 0.49 nm respectively, and the cross Green function $G_{\rho,P}(r)$ vanishes. An increasing coupling increases the range of the Green functions. In Fig. 4, {\it i. e.} $c$=-0.2 \AA, the largest decaying length is equal to 0.59 nm for $G_{\rho}(r)$ and $G_{P}(r)$.  

To conclude this section, we note that the form of coupling between ${\bf P}$ and $\delta\rho$ is imposed by the symmetry of the problem and will always lead to an increase of the response of the system.  The three susceptibilities associated with this model and given in Eqs. (\ref{chirhoc}-\ref{chirhop}) were determined for SPC/E water, in particular $\chi_{\rho,P}(q)$ that was shown to be non vanishing \cite{jeanmairet2016}. This model could be used to propose a tractable analytical theory for water.

\section{Model for water}
The aim of this section is to qualitatively reproduce certain structural properties of water using the present Gaussian field model. Note that water is not a dipolar molecule as it presents three punctual charges, so the  continuous field ${\bf P}(r)$ can be defined as $\nabla\cdot P(r)=\rho_b$, $\rho_b$ the bond charge density.  
We choose the free parameters of this model ($\alpha$, $\kappa_l$, $\nu$, $\lambda$, $c$), so that we can reproduce the positions and values of the maxima of the SPC/E water susceptibilities (see Appendix D for details). Note that $K_{\rho}$ and $K$ are fixed to reproduce macroscopic properties of water.
The results are plotted in Fig. 5.
\begin{figure*}
\includegraphics[scale=0.3]{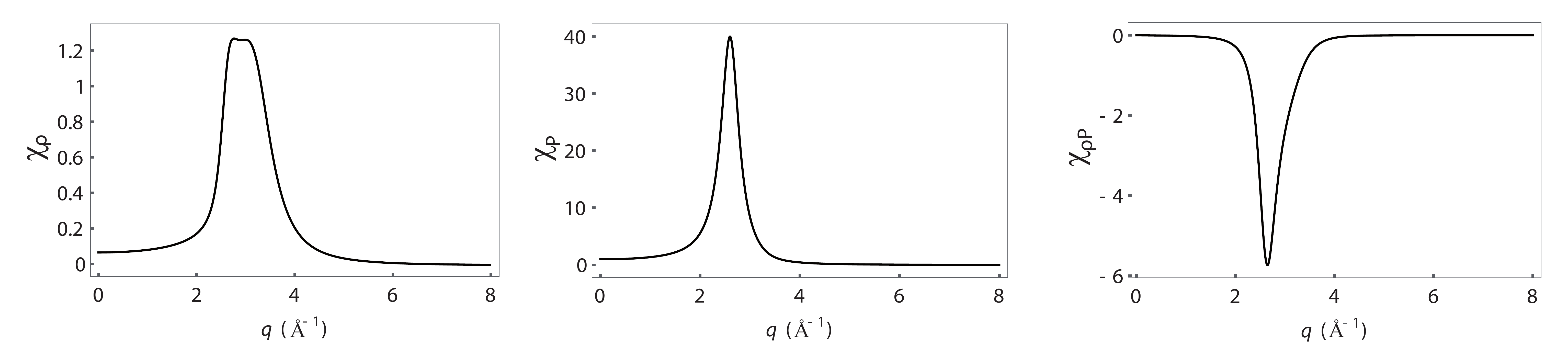}
\caption{Response functions for a model for water. {\bf a.} Density susceptibility $\chi_{\rho}$  (Eq. \ref{chirhoc}) as a function of $q$. {\bf b.} Dielectric susceptibility $\chi_P$ (Eq. \ref{chipc}) as a function of $q$. {\bf c.}  Cross susceptibility $\chi_{\rho,P}$ (Eq. \ref{chirhop}) as a function of $q$. The values of the parameters are given in Eq. (\ref{parameterwater}).}
\end{figure*}
The amplitude, the position and the sign of the three susceptibility maxima are qualitatively comparable to the ones of the susceptibilities of SPC/E water. Moreover, the peak of $\chi_{\rho}(r)$ is broader than in the absence of coupling and associated with two separated maxima, reproducing an experimentally observed aspect of the water structure factor \cite{sellberg2014}. The structure factor $S(q)$ of water can possess two maxima in the zone $2 \AA^{-1} < q <3.5 \AA^{-1}$ the relative positions of which could be related to the presence of two structures in water: the LDL (low density liquid) and the HDL (high density liquid) \cite{gallo2016}. 
The radial distribution corresponding to these response functions is plotted in Figure 6. The dotted line represents the radial distribution obtained without coupling.
\begin{figure}
\includegraphics[scale=0.3]{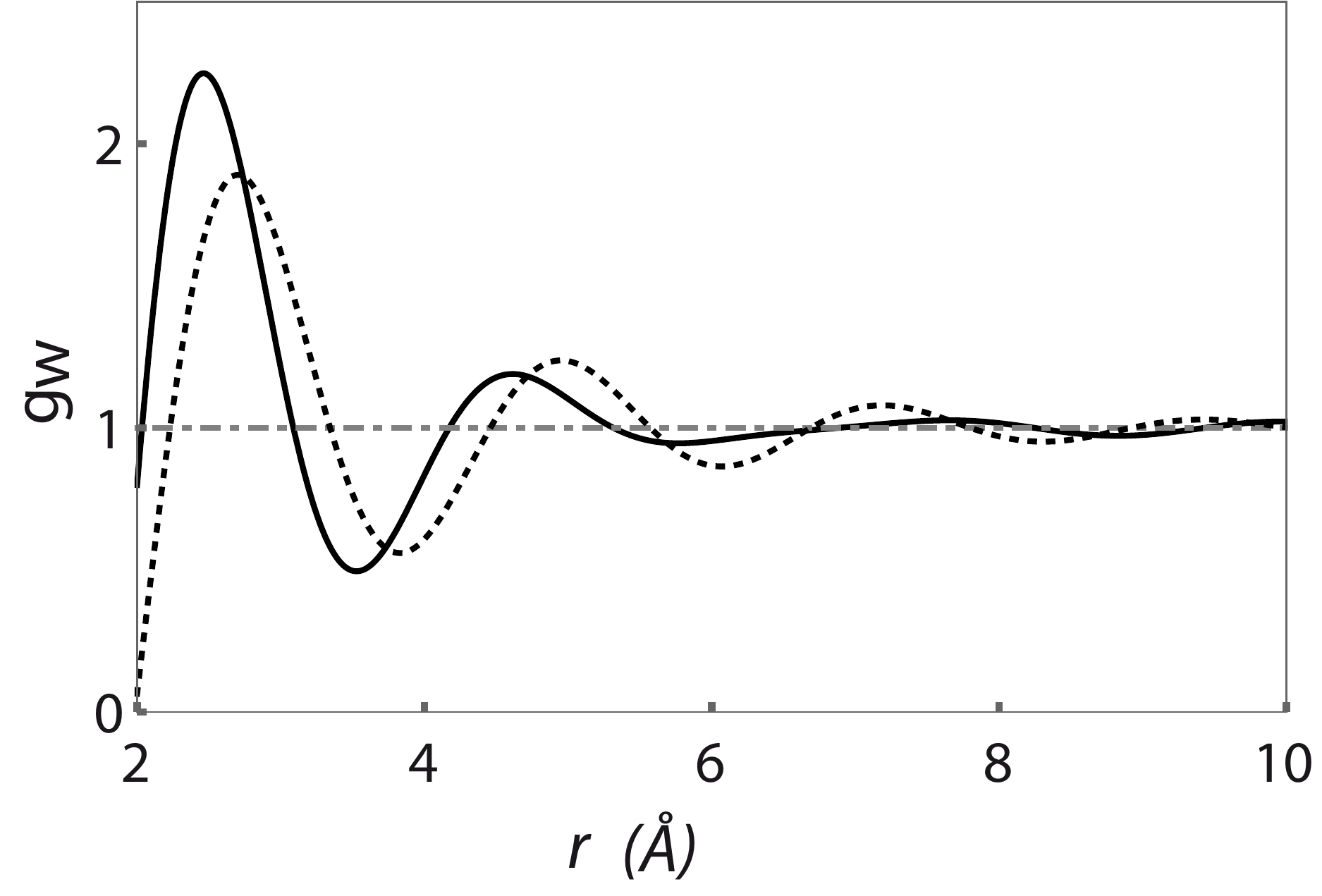}
\caption{Radial distribution function $g_w(r)$ for a model for water. Its expression is given in Eq. (\ref{waterradialdistribution}). The dotted line represent the radial distribution $g(r)$ (Eq. (\ref{raddist})) of the uncoupled model.}
\end{figure}
We observe that the coupling gives rise to a better fit of the numerical and experimental results \cite{sedlmeier2011,sorenson2000}. The first peak located around $r=3\AA$  corresponding to the first solvation shell is more  as observed in data and the correlations decrease rapidly after two oscillations and vanish after 1nm. 

We consider now the response of the medium to an impurity. An atom, located in $r=0$, embedded in the medium can be described as a force field such that the Hamiltonian of the medium is, 
\begin{equation}
\label{Hi}
\mathcal{H}_{part}=\mathcal{H}_{fluid}+\int d^3r\rho(r)\phi_n(r)-\int d^3r{\bf P}(r){\bf E}(r)
\end{equation} 
with $\phi_n(r)$ a Lennard-Jones potential and ${\bf E}(r)$ an electrostatic field,
\begin{eqnarray}
\label{LJ}
\phi_n(r)&=&4\epsilon \left(\frac{\sigma^{12}}{r^{12}}-\frac{\sigma^6}{r^6}\right)\\
\label{ElecField}
{\bf E}(r)&=&-{\rm grad}\phi(r), \quad {\rm with} \quad \phi(r)=\frac{Q}{4\pi\epsilon_0 r},
\end{eqnarray}
where $\epsilon$ and $\sigma$ are Lennard-Jones parameters (using Lorentz-Berthelot mixing rules with Lennard-Jones parameters of the solvent) and $Q$ being the point charge. 

A punctual impurity creates an isotropic response of the solvent given by $\delta\rho(r)$ and $P(r){\bf e_r}$,  which is the sum of two terms: a term generated by the Lennard-Jones potential, $\delta \rho_{n}(r)$ and $P_{n}(r){\bf e_r}$, and a term generated by the electrostatic field, $\delta \rho_{n}(r)$ and $P_{n}(r)$, respectively. 
The density and the polarization response in the direct space, $\delta \rho(r)=\delta \rho_{n}(r)+\delta \rho_{e}(r)$ and $P(r)=P_n(r)+P_e(r)$, are derived by minimizing Eq. (\ref{Hi}). Details of calculations are given in Appendix E.
\begin{figure*}
\includegraphics[scale=0.6]{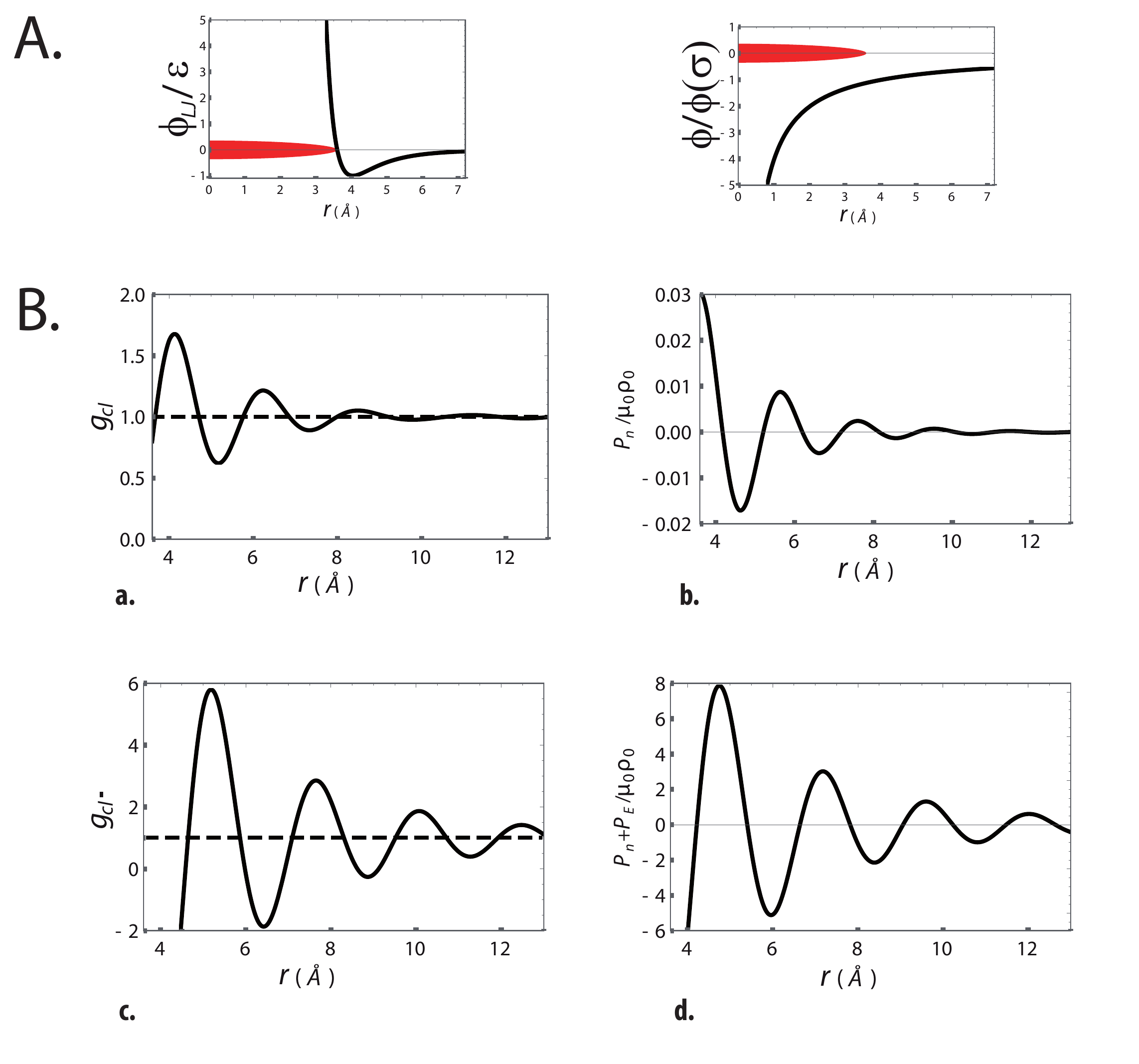}
\caption{A. Representation af the Lennard-Jones potential and electrostatic potential generated by the chlore (red disk). B. a., b. Response of water in density and polarization to the chlore.  c., d. Response of water in density and polarization to the chloride. }
\end{figure*}

Note that we consider only the linear response of the medium to the inclusion without taking into account the nonlinearity coming from the excluded volume effects generated by inclusions embedded in the fluid. This could be done for this model in a second time following the formalism of Li, Kardar and Chandler, which we plan to do in a future publication. 

The response of the medium to a chlore atom ($\sigma_{cl}=3.6$ \AA, $\epsilon_{cl}=9,6.10^{-22}$ J, $Q_{cl}$=0 C) and a chloride ($\sigma_{cl-}=3.6$ \AA, $\epsilon_{cl-}=9,6.10^{-22}$ J, $Q_{cl-}=1,6.10^{-19}$ C) are presented in Fig. 7.
 The results can be compared to molecular dynamics results\cite{jeanmairet2016}.

Figure 7a. represents the radial distribution function $g_{cl}(r)=1+\frac{\delta\rho_n(r)}{\rho}$, the density response of the medium to an atom of chlore, while Figure 7b. represents the distribution of polarization  $ P_n(r)/\mu_0 \rho_0$, the polarization response of the medium to a chlore atom.
The qualitative aspects of the responses are in agreement with the molecular dynamics results. In particular, the amplitude of the polarization response is similar to the molecular dynamics results. 
Figure 7 c. represents the density response $g_{Cl-}=1+\frac{\delta \rho_n(r)+\delta \rho_e(r)}{\rho_0}$ to a chloride and the plot 7. d. is the polarization response $(P_n(r)+P_e(r))/\mu_0\rho_0$. We observe that the order of magnitude of the density and polarization response are in agreement with molecular dynamics\cite{jeanmairet2016}. The response is overestimated in both cases and the decay of the response is too slow, which illustrates that both the dielectric susceptibility $\chi_P(q)$ and the coupling susceptibility $\chi_{\rho,P}(q)$ are sharper than the ones of water.

\section{Discussion and Conclusion}
In this paper, we have introduced a Gaussian field model for polar fluids as a function of two order parameters, the density and the polarization. A Landau-Ginzburg Hamiltonian with a development in $q^4$ for the excess density energy and the dielectric energy is proposed. We showed that a simple linear coupling term whose form is imposed by the symmetry of the problem can give rise to susceptibilities that qualitatively reproduce the susceptibilities of water and consequently its radial distribution of water and the response of the fluid to a perturbation. 

Problems related to liquids and to water in particular are often envisaged using molecular dynamics and data analysis which give access to a large amount of microscopic details and cost an exorbitant computer time. It is interesting to develop in parallel an arsenal of tools to describe water in an analytical framework. We believe that the approach presented here which combines a nonlocal dielectric description of water and an approximate description of density fluctuations is an interesting first step toward a model for water at an intermediate length-scale.
 
\section*{Acknowledgments}
We would like to thank M. Levesque, D. Borgis and N. Dupuis for discussions and revisions.

\appendix
\section{Correlation of polarization without coupling}
The polarization correlations are equal to 
\begin{equation}
\langle P_i(0)P_j(r) \rangle=k_bT\epsilon_0\left(T_0(r)\left(\delta_{ij}-\frac{r_ir_j}{r^2}\right)+L_0(r)\frac{r_ir_j}{r^2}\right),
\end{equation}
where $T_0(r)$ and $L_0(r)$  are the transverse and longitudinal correlation functions. These functions were studied in reference \cite{berthoumieux2015}. The exponential decay length $\lambda_{e,P}$  and the oscillatory decay length $2\pi \lambda_{o,P}$ are equal to 0.49 nm and 0.24 nm respectively. 
The correlation functions of the polarization are equal to: 
\begin{eqnarray}
\label{Lo}
 &&L_0(r)=\frac{e^{-r/\lambda_{e,P}}}{4\pi(K+1)r}\cos\left(\frac{r}{\lambda_{o,P}}\right)\nonumber\\
 & \times &\Big(\frac{2}{r^2}
     \nonumber
     +\frac{1}{r\lambda_{o,P}}\left(R_P+\frac{1}{R_P}\right)\Big)\nonumber
  \\ &+&
        \frac{e^{-r/\lambda_{e,P}}}{4\pi(K+1)r}\sin\left(\frac{r}{\lambda_{o,P}}\right)\Bigg(\frac{1}{\lambda_{o,P}^2}\left(\frac{1}{2R_P^3}+\frac{1}{R_P}+\frac{R_P}{2}\right)\nonumber\\ 
  &+&\frac{1}{r}\left(\frac{1}{\lambda_{o,P}}+\frac{1}{\lambda_{o,P} R_P^2}\right)+\frac{1}{r^2}\left(\frac{1}{R_P}-R_P\right)\Bigg)\nonumber\\
  &-&\frac{1}{2\pi (K+1)r^3}, \nonumber\\
  \label{2ToLo}
&&2T_0(r)+L_0(r)=\frac{(1+R_P^2)^2}{8\pi (1+K) R_P\lambda_{e,P}^2r}e^{-r/\lambda_{e,P}}\sin(\frac{r}{\lambda_{o,P}}), \nonumber
\end{eqnarray}
with $\lambda_{e,P}=\sqrt{2}/q_{0,P}\sqrt{1/\sqrt{\zeta,P}-1}$,
$\lambda_{o,P}=\sqrt{2}/q_{0,P}\sqrt{1/\sqrt{\zeta,P}+1}$,
$R_P=\lambda_{e,P}/\lambda_{o,P}$,
$\zeta_p=\alpha q_{0,P}^4/(1+K)=\left(1+K-1/\chi_P^m
\right)/(1+K)$, $q_{0,P}=\sqrt{-\kappa_l/2\alpha}$, and $\chi_P^m$ the maximum value of $\chi_{P,0}(q)$.
\section{Susceptibility of the system in the presence of coupling}
The kernel $\chi_c^{-1}(q)$ introduced in Eq. (\ref{Hcoupled}) is a 4x4 matrix which is equal to: 
{\tiny
\begin{eqnarray}
\chi_{c}^{-1}(q)= \left(\begin{array}{cccc}\frac{k_bT}{\rho_0}\chi_{\rho,0}^{-1}(q) & ic\frac{k_bT}{\mu_0\rho_0}q_x & ic\frac{k_bT}{\mu_0\rho_0}q_y & ic\frac{k_bT}{\mu_0\rho_0}q_z \\
- ic\frac{k_bT}{\mu_0\rho_0}q_x &\frac{1}{\epsilon_0} \chi_{P,0}^{-1}(q)\frac{q_x^2}{q^2} &\frac{1}{\epsilon_0} \chi_{P,0}^{-1}(q)\frac{q_xq_y}{q^2} &\frac{1}{\epsilon_0} \chi_{P,0}^{-1}(q)\frac{q_xq_z}{q^2} \\
- ic\frac{k_bT}{\mu_0\rho_0}q_y & \frac{1}{\epsilon_0} \chi_{P,0}^{-1}(q)\frac{q_xq_y}{q^2} &\frac{1}{\epsilon_0} \chi_{P,0}^{-1}(q)\frac{q_y^2}{q^2} & \frac{1}{\epsilon_0} \chi_{P,0}^{-1}(q)\frac{q_yq_z}{q^2} \\
- ic\frac{k_bT}{\mu_0\rho_0}q_z & \frac{1}{\epsilon_0} \chi_{P,0}^{-1}(q)\frac{q_xq_z}{q^2} & \frac{1}{\epsilon_0} \chi_{P,0}^{-1}(q)\frac{q_yq_z}{q^2} & \frac{1}{\epsilon_0} \chi_{P,0}^{-1}(q)\frac{q_z^2}{q^2} \\
 \end{array}\right).\nonumber
\end{eqnarray}}%
The corresponding susceptibility is obtained by inverting $\chi_c^{-1}(q)$.

\section{Green functions}

We give the expressions of Green functions of the system described by the Hamiltonian given in Eq. (\ref{Hcoupled}) in real space.
\begin{itemize}
\item  Expression of $G_{\rho}(r)$
\begin{eqnarray}
\label{Grho}
G_{\rho}(r)&=&\frac{1}{(2\pi)^3}\int d^3qe^{i qr\cos(\theta)} \chi_{\rho,c}(q)\\
&=&\frac{1}{4\pi r \alpha\nu}\Sigma_{i=1}^4a_ie^{iq_i r}
\end{eqnarray}
\item Expression of $G_{\rho,P}(r)$
\begin{eqnarray}
\label{Grhop}
G_{\rho,P (i)}(r)&=&\frac{1}{(2\pi)^3}\int d^3qe^{qr\cos(\theta)} \chi_{(\rho,P),i}(q)\\
&=&G_{\rho,P}(r)\frac{{\bf r}}{r}\cdot {\bf e}_i \quad {\rm with} \nonumber \\
 G_{\rho,P}(r)&=&\sqrt{\frac{k_bT \epsilon_0}{\rho_0}}\frac{c}{4\pi \mu_0\alpha\nu}\frac{d}{dr} \left(\Sigma_{i=1}^4 b_i\frac{e^{iq_ir}}{r}\right) 
\end{eqnarray}
where $e_i$ is the unit vector of the orthonormal basis.
\item Expression of the matrix $G_{P}(r)$
\begin{eqnarray}
\label{Gp}
G_{P(ij)}(r)&=&\frac{1}{(2\pi)^3}\int d^3qe^{qr\cos(\theta)} \chi_{P,c}(q)\frac{q_iq_j}{q^2} \\
&=&T_c(r)\left(\delta_{ij}-\frac{r_ir_j}{r^2}\right)+L_c(r)\frac{r_ir_j}{r^2}
\end{eqnarray}
with
\begin{eqnarray}
\label{Lc}
L_c(r)&=&\frac{-1}{4\pi\alpha\nu}\frac{d^2}{dr^2}\left(\Sigma_{i=1}^4 c_i\frac{-1+e^{iq_ir}}{q_i^2 r}\right)\\
\label{TcLc}
2T_c(r)+L_c(r)&=&\frac{1}{4\pi\alpha\nu}\Sigma_{i=1}^4\frac{c_i}{r}e^{iq_ir}
\end{eqnarray}
\end{itemize}
and where $q_i$, $(i=1, ..., 4)$ are the poles of the susceptibility functions, such that 
$(1+K+\kappa_l q^2+\alpha q^4)(K_{\rho}+\lambda q^2+\nu q^4)-c^2q^2\frac{k_bT\epsilon_0}{\mu_0^2\rho_0}=\alpha\nu\Pi_i(q^2-q_i^2)$. Note that $q_i$ are associated with a positive imaginary part.
The coefficients $(a_i, b_i, c_i)$, $(i=1,...,4)$ obey the following equations
{\small \begin{eqnarray}
\label{coeffeq}
{\bf M}\cdot{\bf a}=\left(\begin{array}{c} 0\\ \alpha \\ \kappa_l \\ 1+K \end{array}\right),  {\bf M}\cdot{\bf b}=\left(\begin{array}{c} 0\\ 0 \\ 0 \\ 1 \end{array}\right),  {\bf M}\cdot{\bf c}=\left(\begin{array}{c} 0\\ \nu \\ \lambda \\ K_{\rho} \end{array}\right)
\end{eqnarray} }%
and the matrix M is equal to 
\begin{widetext}
\begin{eqnarray}
\label{M}
M=\left(\begin{array}{cccc} 1 & 1 & 1 & 1 \\
-(q_2^2+q_3^2+q_4^2) & -(q_1^2+q_3^2+q_4^2) & -(q_1^2+q_2^2+q_4^2) & -(q_1^2+q_2^2+q_3^2) \\
q_2^2q_3^2+q_2^2q_4^2+q_3^2q_4^2 & q_1^2q_3^2+q_1^2q_4^2+q_3^2q_4^2 & q_1^2q_2^2+q_2^2q_4^2+q_1^2q_4^2 & q_1^2q_2^2+q_2^2q_3^2+q_3^2q_1^2 \\
-q_2^2q_3^2q_4^2 & -q_1^2q_3^2q_4^2 & -q_1^2q_2^2q_4^2 &-q_1^2q_2^2q_3^2 \end{array}\right).
\end{eqnarray}
\end{widetext}
In the absence of coupling one finds for the Green functions of the system:
\begin{itemize}
\item $G_{\rho}(r)=\langle \delta \rho (r) \delta \rho (0) \rangle_0/\rho_0$, with $\langle \delta \rho (r) \delta \rho (0) \rangle_0$ given in Eq. (\ref{rhorho0}).
\item $G_{\rho,P}(r)= 0$
\item $G_{P (ij)}(r)=T_0(r)\left(\delta_{ij}-\frac{r_ir_j}{r^2}\right)+L_0(r)\frac{r_ir_j}{r^2}$ with $T_0(r)$ and $L_0(r)$ given in Appendix A. 
\end{itemize}

 The coupling modifies the range of the correlations. The Green functions $G_{\rho}(r)$,  $G_{\rho, P}(r)$, $L_c(r)$ and $T_c(r)$ are the sum of two oscillating functions in decaying envelopes of range $\lambda_{e,1}$ and $\lambda_{e,2}$  with $\lambda_{e,1}<\lambda_{e,2}$ and $\lambda_{e,1} \rightarrow \lambda_{e,\rho}$ and $\lambda_{e,2}\rightarrow \lambda_{e,P}$ for $|c| \rightarrow 0$. 
As presented in Figure A. 1, $\lambda_{e,2}$ increases with an increase of coupling strength $|c|$, whereas $\lambda_{e,1}$ slightly decreases. 
The relative weight of each decaying envelope varies with the strength of the coupling.
Its is presented in Fig A. 2, A.3 and A. 4  
\begin{itemize}
\item $G_{\rho}(r)$: the relative contribution of the envelope decaying in $\lambda_{e,2}$ compared to the one decaying in $\lambda_{e,1}$ increases with an increasing coupling strength and vanishes in the absence of coupling. The function decaying in $\lambda_{e,1}$ remains practically unchanged as the coupling strength increases. (See Fig A. 2.) 
\item $G_{\rho,P}(r)$: the cross Green function vanishes in the absence of coupling. For increasing coupling, the function is dominated by the contribution decaying in $\lambda_{e,2}$. (See Fig. A. 3.)
 \item $G_{P(i,j)}(r)$ : the functions $L_c(r)$ and $2T_c(r)+L_c(r)$ characterizing the response in polarization are a sum of envelopes decaying in $e^{-\lambda_{e,i}}/r^j$, with $(i=1,2)$ and $(j=1,... ,3)$. Similarly to the density and cross Green functions, an increasing coupling strength enhances the range of the response by increasing the weight of the functions decaying in $\lambda_2$.  In the absence of coupling, the terms decaying in $\lambda_{e,1}$ vanish. 
\end{itemize}
\begin{figure*}
\includegraphics[scale=0.3]{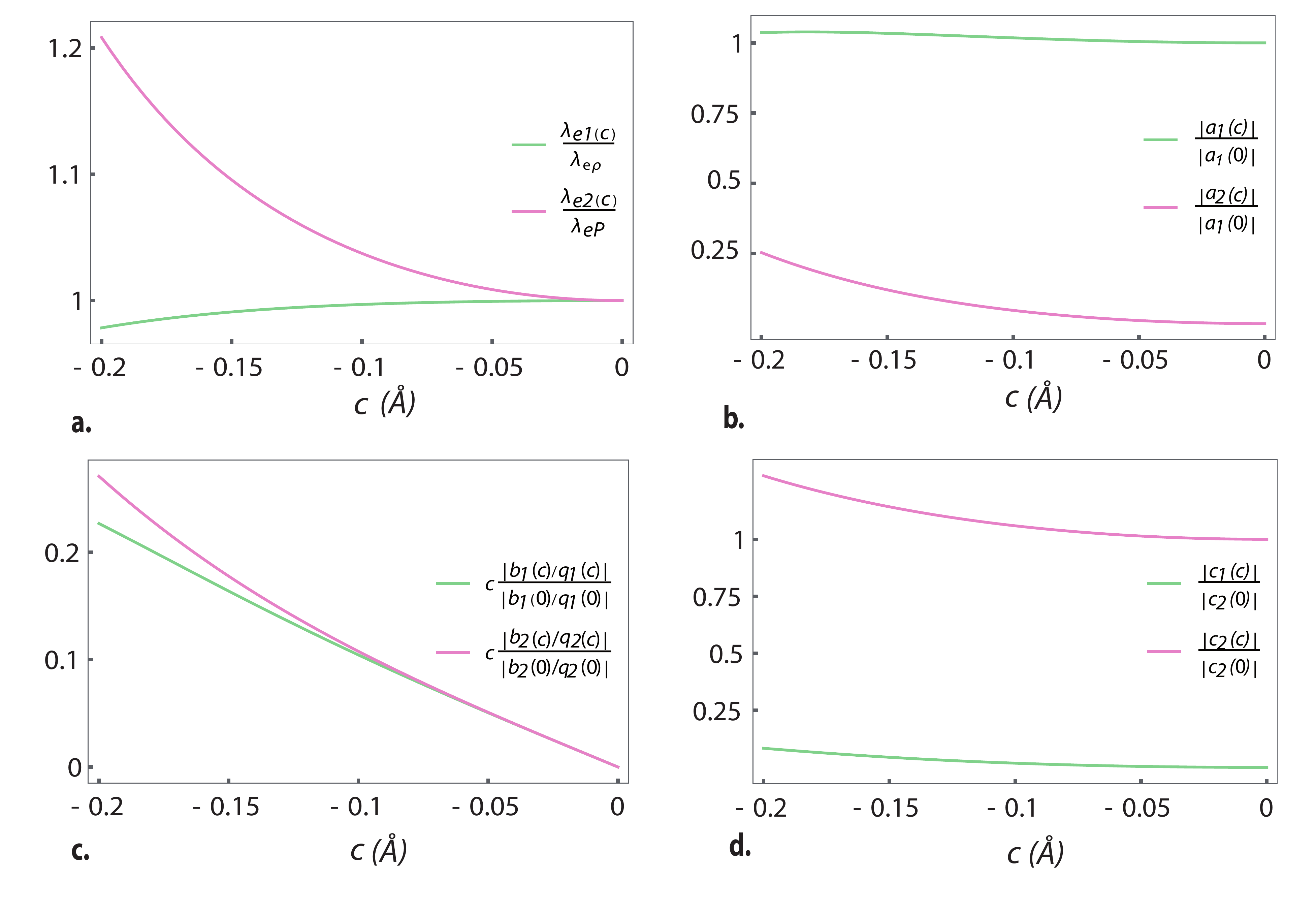}
\caption{Influence of the coupling strength on the range of the Green functions. {\bf a}. Normalized decaying lengths $\lambda_{e_1}/\lambda_{e\rho}$ and $\lambda_{e_2}/\lambda_{eP}$ as a function of $c$. {\bf b}. Normalized weight of the functions decaying in $\lambda_{e_1}$ and $\lambda_{e_2}$ in $G_{\rho}(r)$ as a function of $c$. {\bf c}.  Normalized weight of the functions decaying in $\lambda_{e_1}$ and $\lambda_{e_2}$ in $G_{\rho,P}(r)$ as a function of $c$. {\bf d}.  Normalized weight of the functions decaying in $\lambda_{e_1}$ and $\lambda_{e_2}$ in $2T_c(r)+L_c(r)$ as a function of $c$.}
\end{figure*}
\section{Parameter values for a description of water}
In this appendix, we determine the values of the parameters $\alpha$, $\kappa_l$, $\lambda$, $\nu$ and $c$ that can reproduce certain features of the response functions of water.
We choose 5 conditions for the susceptibilities and its derivatives that can be written as follows
\begin{eqnarray}
\chi_{P}(2.6)&=&40, \quad \frac{\partial \chi_P(2.6) }{\partial q}=0, \nonumber\\
\chi_{\rho}(3)&=&1.25, \quad \frac{\partial \chi_{\rho}(3) }{\partial q}=0, \quad \chi_{\rho,P}(2.6)=6.5. \nonumber
\end{eqnarray}
The aforementioned equations are functions of ($\alpha$, $\kappa_l$, $\lambda$, $\nu$, $c$). As a reminder, the value of the macroscopic properties of the medium is fixed in this paper to $K_{\rho}=14.4$ and $K=1/70$, the values obtained for SPC/E water \cite{kusalik1994}.
Solving these equations one obtains the following values
\begin{eqnarray}
\label{parameterwater}
\alpha_w&=&0.024 \AA^{-4}, \quad \kappa_{lw}=-0.3\AA^{-2},\quad \nu_w=0.12 \AA^{-4}, \nonumber\\
\lambda_w&=&-2.5\AA^{-2}, \quad c_w= -0.90 \AA
\end{eqnarray} 
and the corresponding response functions are plotted in Figure 5. 

Note that the goal is not to derive the set of parameters that gives the "best" susceptibilities to reproduce experimental results but to catch some of their important aspects. In this frame, the plots presented in the main text are satisfying because the three susceptibilities are qualitatively similar to the numerical measures.

The radial distribution function associated to this set of parameter is written as 
\begin{equation}
\label{waterradialdistribution}
g_w(r)=1+ G_{\rho}(r)/\rho_0
\end{equation}
where $G_{\rho}(r)$ is the Green function given in Eq. (\ref{Grho}) with the coefficients $a_i$ and the poles $q_i$, ($i=1,...,4$ ) associated with the parameters ($\alpha_w$, ..., $c_w$) previously given.

\section{Density and polarization response}
We calculate the response of the medium to an external potential given by Eq. (\ref{Hi}). The potential is radial and will induce a radial polarization $P(r){\bf e }_r$ in the medium that can be written $P(r){\bf e }_r=\nabla \psi(r)$, with $\psi(r)$ a scalar potential. 
The response of the fluid density and polarization is obtained by minimizing the Hamiltonian given in Eq. (\ref{Hi}) that can be written in Fourier space as a function of $\delta\rho(q)$ and $\psi(q)$,
\begin{eqnarray}
\mathcal{H}[\delta \rho,\psi]&=&\frac{1}{2}\int d^3q  \left(\begin{array}{cc} \delta\rho(q) & \psi(q) \end{array} \right)\chi_{c,\psi}^{-1}(q) \left(\begin{array}{c}\delta\rho (-q) \\ \psi(-q) \end{array} \right)\nonumber\\
 &+& \int d^3q \left(\begin{array}{cc} \delta\rho(q) & \psi(q) \end{array} \right)\left(\begin{array}{c}\phi_n^a (-q) \\ q^2\phi(-q) \end{array} \right),
\end{eqnarray}
with 
\begin{eqnarray}
\chi_{c \psi}^{-1}(q)= \left(\begin{array}{cc} \frac{k_bT}{\rho_0}\chi_{\rho,0}^{-1}(q) & c\frac{k_bT}{\mu_0\rho_0}q^2 \\ c\frac{k_bT}{\mu_0\rho_0}q^2 & \frac{q^2}{\epsilon_0}\chi_{P,0}^{-1}(q) \end{array} \right)
\end{eqnarray}
The Lennard-Jones potential $\phi(r)$ defined in Eq. (\ref{LJ}) does not have a Fourier Transform due to its high divergence in 0. We approximate it by the potential $\phi^a (r)$ defined as follow,
\begin{equation}
\phi_n^a(r)=\left \{ \begin{array}{ll}
\ 20\epsilon \quad {\rm for } \quad r<\sigma \\
4\epsilon\left(\frac{\sigma^{12}}{r^{12}}-\frac{\sigma^6}{r^6}\right) \quad {\rm for} \quad r>\sigma
\end{array}\right.
\end{equation}
which Fourier transform is numerically determined. 
The response of the medium to the inclusion is given by a minimization of the Hamiltonian with respect to $\delta\rho(q)$ and $\psi(q)$ and one finds,
{\small \begin{eqnarray}
\left(\begin{array}{c} \delta \rho (q) \\ \psi(q) \end{array} \right) &=& -\chi_{c, \psi} (q) \left(\begin{array}{c} \phi_n (q) \\ q^2 \psi(q) \end{array} \right)\\
&=&\frac{1}{\chi_{P,0}^{-1}(q)\chi^{-1}_{\rho,0}(q)-c^2q^2\frac{k_bT\epsilon_0}{\mu_0^2 \rho_0}}\\
&\times &\left(\begin{array}{cc} -\frac{\rho_0}{k_bT}\chi_{P,0}^{-1}\phi^a_n(q) &  \frac{c\epsilon_0}{\mu_0}q^2\phi(q) \\ \frac{c \epsilon_0}{\mu_0}\phi^a_n(q) & -\epsilon_0\chi_{\rho,0}^{-1}(q)\phi(q) \end{array} \right)
\end{eqnarray}}%

The density and the polarization are the sum of two contributions, one triggered by the the Lennard-Jones potential and the other one by the electrostatic potential.
The first one is derived numerically,
\begin{eqnarray}
\delta\rho_n(r)&=&-\frac{1}{(2\pi)^3}\frac{\rho_0}{k_bT}\int d^3q \frac{\chi_{P,0}^{-1}(q)\phi_n^a(q)e^{iqr}}{\chi_{P,0}^{-1}(q)\chi_{\rho,0}^{-1}(q)-c^2q^2\frac{k_bT\epsilon_0}{\mu_0^2 \rho_0}} \\
\psi_n(r)&=&\frac{1}{(2\pi)^3}\frac{c \epsilon_0}{\mu_0}\int d^3q \frac{\phi_n^a(q)e^{iqr}}{\chi_{P,0}^{-1}(q)\chi_{\rho,0}^{-1}(q)-c^2q^2\frac{k_bT\epsilon_0}{\mu_0^2 \rho_0}}.
\end{eqnarray}
 The electrostatic contribution can be expressed analytically in real space by writing $\phi(q)=\frac{Q}{\epsilon_0 q^2}$. It gives, 
\begin{eqnarray}
\delta \rho_{el}(r) &=& \frac{cQ}{4\mu_0\pi r\nu \alpha}\Sigma_{i=1}^4b_ie^{iq_i r}, \\
\psi_{el} (r) &=& -\frac{Q}{4\pi r \alpha \nu} \Sigma_{i=1}^4c_i\frac{-1+e^{iq_ir}}{q_i^2}.
\end{eqnarray}
The polarization ${\bf P}(r)=P(r){\bf e_r}$ is given by the relation $P(r)=\psi'(r)$.

\bibliographystyle{unsrt}
\bibliography{bibcouplage}
\end{document}